\documentclass[11pt]{article}
\usepackage{fullpage}
\usepackage{booktabs}
\usepackage[utf8]{inputenc}
\usepackage{lineno, graphicx, amsmath, amssymb, tabularx}
\usepackage{titlesec}
\usepackage[round,authoryear,sort]{natbib}
\usepackage[english]{babel}
\usepackage{subcaption}
\usepackage{cancel}
\usepackage{amsmath}
\usepackage{url}
\usepackage{cleveref}
\usepackage{geometry}
\usepackage{comment}
\usepackage{array}
\usepackage{xcolor}
\usepackage{setspace}
\usepackage{xr-hyper}
\newcolumntype{L}[1]{>{\centering\arraybackslash}p{#1}}
\usepackage{float} %
\usepackage{lineno} %
\usepackage{graphicx} %
\usepackage{newpxtext,newpxmath,paralist}

\usepackage{physics}
\usepackage{setspace}
\linenumbers
\title{\textbf{Disentangling the interactive effects of anthropogenic disturbances on biodiversity}}
\author{Isaac Planas-Sitjà*$^{1, 2}$, Ryosuke Iritani $^1$ \& Adam L. Cronin$^2$}
\date{}

\begin{document}
\nolinenumbers
\maketitle
\doublespacing

{\parindent0pt
	
* correspondent author
\\
$^1$ RIKEN Wako Campus, Division of Fundamental Mathematical Science, 2-1 Hirosawa, Wako, Saitama 351-0198 Japan

$^2$ Tokyo Metropolitan University, Department of Biological Sciences, 1-1 Minami-Osawa, Tokyo, Japan 
\\
\\
\\
\\
\\
\\
\\

\pagebreak

\section{Abstract} %

Anthropogenic activity threatens biodiversity through climate change, habitat fragmentation, and increasing frequency and scale of disturbance. Various theoretical studies have sought to shed light on how these factors could promote or hinder the coexistence of species. However, our understanding of the relative importance of, and interactions between, these factors remains limited. In this study, we employ a theoretical approach integrating three commonly cited coexistence mechanisms -- the competition-colonisation trade-off, the intermediate disturbance hypothesis, and spatial heterogeneity -- into a unified model. We implement a novel method to integrate habitat autocorrelation into a system of differential equations, to create a simple and flexible model that can be used to investigate coexistence of multiple species arranged in a competitive hierarchy under different disturbance and habitat structure scenarios. Using this model, we find that considering interactions between different mechanisms is crucial for explaining the coexistence of species. Biodiversity patterns alternative to the uni-peak curve predicted by the intermediate disturbance hypothesis (e.g., bimodal) emerge along disturbance gradients as habitat fragmentation increases. Furthermore, habitat loss outweighs habitat autocorrelation effects in highly disturbed scenarios, yet autocorrelation can shape species coexistence under low disturbance. These findings underscore the need to integrate spatial and temporal mechanisms in biodiversity management.\par
\

\textbf{Key words:} Biodiversity, competition-colonisation trade-off, intermediate disturbance hypothesis, habitat fragmentation, dispersal, metacommunity

\pagebreak
\section{Introduction}

Human activity threatens natural ecosystems directly through transformation of the landscape and increasing temporal disturbances, and indirectly through climate change \citep{Foley2005}. These impacts influence population dynamics, distribution of species, interactions between species, and ultimately the maintenance of biodiversity \citep{Heino2013}. Disentangling the complex array of human impacts poses a challenge to scientists attempting to develop our understanding of the mechanisms sustaining biodiversity \citep{Duputie2013, Heino2013, Cote2017}. Metacommunity theory has proved valuable in addressing this challenge by explaining the conditions that support coexistence of species, providing valuable insights into the maintenance of biodiversity. It is thus not surprising that this problem has generated a broad spectrum of metacommunity models, which could be broadly divided into four paradigms: neutral theory, species sorting, mass effects and patch dynamics \citep{Shoemaker2016, Chesson2000}. While not mutually exclusive, these paradigms tend to emphasize either the effect of non-spatial mechanisms (i.e., differences in species traits) or spatial mechanisms (i.e., niche differences or disturbances) \citep{Shoemaker2016, Adler2007, Chesson2000, Bolker2003, Snyder2003}. By non-spatial mechanisms we refer to those that act in the absence of any spatial heterogeneity, and are often described as equalizing mechanisms because they reduce fitness differences among species. In comparison, spatial mechanisms, mostly referred to as stabilising mechanisms, tend to limit species fitness by inducing intraspecific negative effects and often depend on environmental fluctuations in time or space (see \citet{Bolker2003, Chesson2000, Adler2007} for details).\par 
\

Metacommunity models have helped understand the role of spatial and non-spatial mechanisms in community structure and species coexistence \citep{Whittaker1962, Leibold2004, Hubbell2001, Levins1968, Shmida1985, Chesson1994, Connell1978, Roxburgh2004}. For instance, the competition-colonisation trade-off (CCTO), a non-spatial mechanism, assumes that colonisation ability comes at a cost of competitive ability \citep{Skellam1951, Levins1971, Hastings1980, Tilman1994}, and explains how species arranged in a competitive hierarchy could coexist: colonisers escape competitive exclusion by dispersing to newly available patches while competitors dominate stable patches \citep{Yu2001, Kisdi2003, Gross2008, Calcagno2006, Liao2022}. This theoretical framework has thereby become one of the most broadly investigated non-spatial mechanisms in species coexistence, and empirical studies show that the CCTO can play a relevant role in the assembly of plant and microbial communities \citep{Wetherington2022, Levine2002, Moles2006, Yu2004}. Models introducing spatial structure mainly focus on the influence of habitat fragmentation on biodiversity \citep{Leibold2004, Leibold2015}, and predict that dispersal ability should be selected for in landscapes with low fragmentation, but selected against under high fragmentation due to the high cost of moving across a fragmented landscape \citep{Olivieri1995, Travis1999a, Bonte2010, Cote2017}. In support of this theory, \citet{Bonte2006} observed that tiptoe behaviour (a precursor of aerial dispersal) in the spider \textit{Pardosa monticola} decreased with habitat fragmentation, and two butterfly species exhibited decreased thoracic muscular mass (needed for dispersal) with increasing habitat isolation \citep{Dempster1991}. An alternative and frequently cited spatial mechanism for species coexistence is the intermediate disturbance hypothesis (IDH) \citep{Wilson1990}. This theory predicts a unimodal disturbance-diversity relationship with a biodiversity peak at intermediate levels of temporal disturbance, as this facilitates persistence of coloniser species facing competitive exclusion, while high disturbance levels are detrimental to all species \citep{Hastings1980, Liao2022, Roxburgh2004}. Various studies have also found support for these predictions. For instance, \citet{Sousa1979} investigated the effects of disturbance on a marine intertidal community, and found that communities on boulders subjected to intermediate-frequency disturbance were more diverse than those on undisturbed or frequently-disturbed boulders.\par 
\

However, models have also failed to predict natural patterns in other cases. For example, there is considerable debate over whether the strict competitive hierarchies at the core of the CCTO are well represented in nature \citep{Adler2000, Yu2001, Kisdi2003, Calcagno2006, Gross2008, Levine2017}. Similarly, questions surround the general applicability of the IDH, as it fails to capture the full complexity of diversity-disturbance patterns observed in nature, such as monotonically declining/increasing or U-shaped \citep{Mackey2001, Cadotte2007, RandallHughes2007, Liao2022}. These criticisms highlight the limitations of some modelling approaches for predicting the complex array of interactive components in natural systems. For example, the difficulty of incorporating spatial processes, such as resource heterogeneity, in mathematical models \citep{Cohen1991, Chesson2000,  Mathias2001b, Massol2010, Leibold2015, Parvinen2020, Parvinen2023}, has led several authors to use computer simulations, which can more easily incorporate spatial components \citep{King2002, Bonte2010, Cronin2016, Planas-Sitja2023}. These spatially explicit simulation studies have often further complicated the issue by contradicting previous mathematical approaches though, showing, for instance, that increased habitat loss in fragmented habitats could select for a high colonisation ability (i.e., higher dispersal) because of the benefits of colonising new habitats and escaping local extinction \citep{Heino2001}. Nonetheless, these predictions have also found empirical support, for example in damselflies \citep{Taylor1995} and butterflies \citep{Schtickzelle2006}.\par 
\

A key factor missing in various studies, and which may be able to shed light on the limitations of the metacommunity models outlined above, is the interactive effects among various mechanisms influencing coexistence of species \citep{Heino2013, Duputie2013}. Because of the intricate nature of modelling natural ecosystems, most studies tend to focus on a single mechanism supporting coexistence, while keeping other factors constant. For example, considering different levels of disparity in competition and colonization ability in a constant environment \citep{Calcagno2006}, or varying disturbances in homogeneous landscapes \citep{Liao2022}, or incorporating landscape heterogeneity with a constant disturbance regime \citep{Cohen1991, Planas-Sitja2023, Bonte2010, Mathias2001b}. Where efforts have been made to integrate multiple components influencing coexistence of species, more subtle patterns have become evident \citep{Shoemaker2016, Snyder2004}. This work suggests that the CCTO, albeit remaining debated for its rigidity in assuming displacement competition, may hold untapped potential for explaining non-linear patterns of biodiversity in response to environmental gradients. Specifically, integrating spatial mechanisms with the CCTO \citep{Gross2008, Kondoh2001, Bolker2003, DosSantos2011} or relaxing its absolute displacement assumption \citep{Calcagno2006, Li2020, Liao2022} can significantly enhance our understanding of the mechanisms promoting biodiversity, offering nuanced insights that were previously inaccessible. For example, the effects of the CCTO can be modulated by disturbance level under certain conditions \citep{Cai2022}, such that high levels of disturbance can boost coexistence of species only when there is heterogeneity in resource availability \citep{Gross2008, Cronin2016}. In addition, the influence of the CCTO can be enhanced when considering the effects of fecundity and dispersal separately, rather than combining these in a single measure of colonisation ability (i.e., number of offspring dispersing outside the site boundaries) \citep{Bolker2003a}. While many of the existing metacommunity models boast complex structures, they often fall short in practical utility for field ecologists who require more streamlined and straightforward tools to bridge theoretical knowledge and empirical observations. As such, simple yet effective models are essential to fostering a meaningful connection between ecological theory and real-world application. Novel integrative approaches with multi-dimensional trade-off scenarios are thus needed to help identify key areas for prioritization of conservation actions \citep{Clark2007, Travis2012, Heino2013}. Along these lines, \citet{Zhang2023} introduced a simple model demonstrating that even if biodiversity tends to decrease with fragmentation, the CCTO can give place to a non-linear oscillatory relationship between fragmentation and biodiversity. These findings altogether highlight how interactions between mechanisms can influence model predictions, and may help explain complex natural patterns \citep{Schtickzelle2006, Cai2022}. By refining the CCTO framework with spatial dynamics and emphasizing simplicity, we can develop a model that not only aligns closely with field realities \citep{Riva2022} but also enhances our predictive capabilities in diverse ecosystems \citep{Zhang2023, Kondoh2001}.\par
\

In this study, we explore the relative importance of, and interactions between, non-spatial and spatial mechanisms of species coexistence on biodiversity. We do so by using a new method to integrate spatial autocorrelation into a system of differential equations. We begin with Tilman's CCTO equation to model species interactions \citep{Tilman1994}, and add to this temporal and spatial anthropogenic effects, including habitat loss, habitat autocorrelation and disturbance. We incorporate these factors into a spatially implicit model, enabling us to disentangle the influences of each factor on biodiversity, understood here as the coexistence of multiple species, and assess the role of interactive effects between these factors. This simple yet general model balances tractability and applicability, making it useful for empirical ecologists who can easily parametrise it with field data, without requiring any specialized mathematical expertise.\par 
\

\section{Methods}

\subsection{General description of the model}
We present a simple and versatile model to describe a metacommunity structured in an unspecified number of habitable patches connected by dispersal events, and subject to random disturbances resulting in local extinctions. This spatially implicit model describes the between-patch dynamics for any number of species differing in their reproductive strategies, in landscapes with varying levels of anthropogenic disturbance (fragmentation and local extinction rate). We assume that only one individual can occupy a given area unit (patch), which could represent a defended territory, or an area occupied by a sessile organism. We use this model to investigate how anthropogenic disturbance influences the degree to which species can coexist, and by extension, its impact on biodiversity.

\subsection{Modelling the dynamics for two species:}

We consider a metacommunity composed of two species differing in their reproductive strategies: a coloniser ($L$) and a competitor ($M$). Each species is defined by its colonisation rate ($c$) and dispersal distance ($d$). The colonisation rate $c$ defines the frequency of reproduction (or fecundity), while the dispersal distance determines the distance the offspring can disperse (1 being global dispersal and 0 being highly restricted local dispersal). We assume the existence of a CCTO, which conforms to the intuition that the colonisation rate and competitive ability are negatively related. Thus, $c$ defines an axis of reproductive investment, on which one extreme might correspond to a more competitive, $k$ strategist species, which grows slower and reproduces less often (or produces fewer offspring), while the other extreme corresponds to a less competitive species, $r$ strategist species, which grows faster and reproduces more often (or produces more offspring). For simplification, we consider that species form a competitive hierarchy \citep{Tilman1994}, where the best competitor always displaces a resident coloniser species. In such a case, and assuming that the mortality ($\mu$) is the same for both species, the coloniser ($dL$) and competitor ($dM$) populations in a homogeneous environment are given by:

\begin{eqnarray}\
	&& \frac{dL}{dt} = \underbrace{L c_L \left(1-L-M \right)}_{\text{Colonised patches}} - \underbrace{M L c_M}_{\text{Competition}} - \underbrace{\mu L}_{\text{Mortality}}\label{eq:1}\\
	&&\frac{dM}{dt} = \underbrace{M c_M \left(1-M\right)}_{\text{Colonised patches}}-\underbrace{\mu M}_{\text{Mortality}} \label{eq:2}\
\end{eqnarray}

The population of the coloniser species will grow through colonisation of empty patches (first term of \eqref{eq:1}), but decline due to the loss of patches through competition and stochastic mortality (second and third terms respectively). Similarly, a competitor species \eqref{eq:2} will acquire new patches by colonisation (empty patches or those occupied by the coloniser) but lose them due to mortality (see \citet{Tilman1994} for more details). Below, we modify these equations to consider differences in landscape heterogeneity: proportion of inhabitable patches and their distribution (see table 1 for description of parameters).

\begin{table}[H]
	\centering
	\caption{Summary of parameters used in the model with 2 species.}
	\label{t:1}
	\begin{tabular}{| L{2cm} | L{10cm} | L{3cm} |}
		\hline
		Parameter & Description & Value \\
		\hline
		& Environmental variables & \\
		\hline
		$H$ & Habitat loss, or proportion of poor patches & [0 - 1] \\
		$A$ & Habitat autocorrelation, from low (0) to high (1) & [0 - 1] \\
		$\mu$ & Mortality due to natural and antrhopogenic disturbances & [0.01, 0.05, 0.1] \\
		\hline
		& Strategies & \\
		\hline
		$c_L$ & Colonisation ability of coloniser strategy & 0.9  \\
		$c_M$ & Colonisation ability of competitor strategy & 0.2  \\
		$d_L$ & Dispersal distance for coloniser strategy & 0.9  \\
		$d_M$ & Dispersal distance for competitor strategy & 0.2  \\
		\hline
	\end{tabular}
\end{table}

\subsubsection{Introducing spatial heterogeneity}

The major consequences of habitat fragmentation are a) the loss of suitable habitat, and b) the splitting up of continuous habitat into smaller and more isolated patches, decreasing the connectivity between habitats \citep{Schtickzelle2006, Cote2017, Zhang2023, Fahrig2017}. Separating these factors has proved challenging in spatially implicit models and they are often treated together, though their effects may differ. Here, we provide a novel method to include spatial autocorrelation in a spatial niche partitioning model, allowing us to consider the effects of habitat loss and patch spatial distribution separately. First, environmental heterogeneity is modelled by considering patches of two qualities: rich  ($R$) and poor ($P$), with rich patches being inhabitable and poor patches being uninhabitable. For simplicity, we fix the carrying capacity of rich patches ($K_R$) to 1, while $K_P$ = 0 (uninhabitable). To simulate habitat loss, we use the parameter $H$, which represents the proportion of poor patches [0 - 1], with the proportion of rich patches correspondingly being (1 - $H$). To simulate the splitting up of the landscape, we use the spatial autocorrelation index ($A$). At one extreme we have continuous zones of the same habitat, represented by high spatial autocorrelation (1), while at the other extreme we have this habitat fragmented in a high number of small patches, reflected by low spatial autocorrelation (0) (\cref{fig:1}A, B). For instance, in the case where $H = 0.5$ and $A = 1$, this corresponds to a landscape with high spatial autocorrelation and with 50\% inhabitable space, effectively generating habitat islands separated by uninhabitable space. Increasing the value of $A$, therefore, results in competitors (non-dispersers) sampling the same environment more often than colonisers (dispersers) \citep{Cohen1991, Planas-Sitja2023}. Second, the mortality rate ($\mu$) is considered to be the result of natural and human actions that vary in the magnitude and/or frequency of disturbance. Because we consider disturbance during a generation (e.g., one year), a high frequency of weak disturbances or low frequency of strong disturbances would amount to the same effect in our model. Assuming a strict competitive hierarchy as above, and that $K_P$ = 0, equations \eqref{eq:1} and \eqref{eq:2} can be written as:\par 
\

\begin{eqnarray}\
	&& \frac{dL_R}{dt} = \underbrace{L_R c_L \lambda_{RL} \left(1-H-L_R-M_R \right)}_{\text{Colonised patches}} - \underbrace{M_R L_R c_M \lambda_{RM}}_{\text{Competition}} - \underbrace{\mu L_R}_{\text{Mortality}}\label{eq:3}\\
	&&\frac{dM_R}{dt} = \underbrace{M_R c_M \lambda_{RM} \left(1-H-M_R\right)}_{\text{Colonised patches}}-\underbrace{\mu M_R}_{\text{Mortality}} \label{eq:4}\
\end{eqnarray}

with $L_R$ ($M_R$) being the coloniser (competitor) population in rich patches. Equations \eqref{eq:3} and \eqref{eq:4} determine the proportion of rich patches occupied by the coloniser and competitor species respectively. For equation \eqref{eq:3}, the first term corresponds to the proportion of rich patches colonised by offspring from existing coloniser individuals. The second and third terms are the patches lost through competition with the offspring of the  competitor species, and mortality, respectively. For equation \eqref{eq:4}, the first term is the proportion of rich patches colonised by offspring of existing competitor individuals, while the second term is the patches lost due to mortality (see qualitative results in \cref{fig:1}D, E).\par 
\

\begin{figure}[H]
	\includegraphics[scale=0.82]{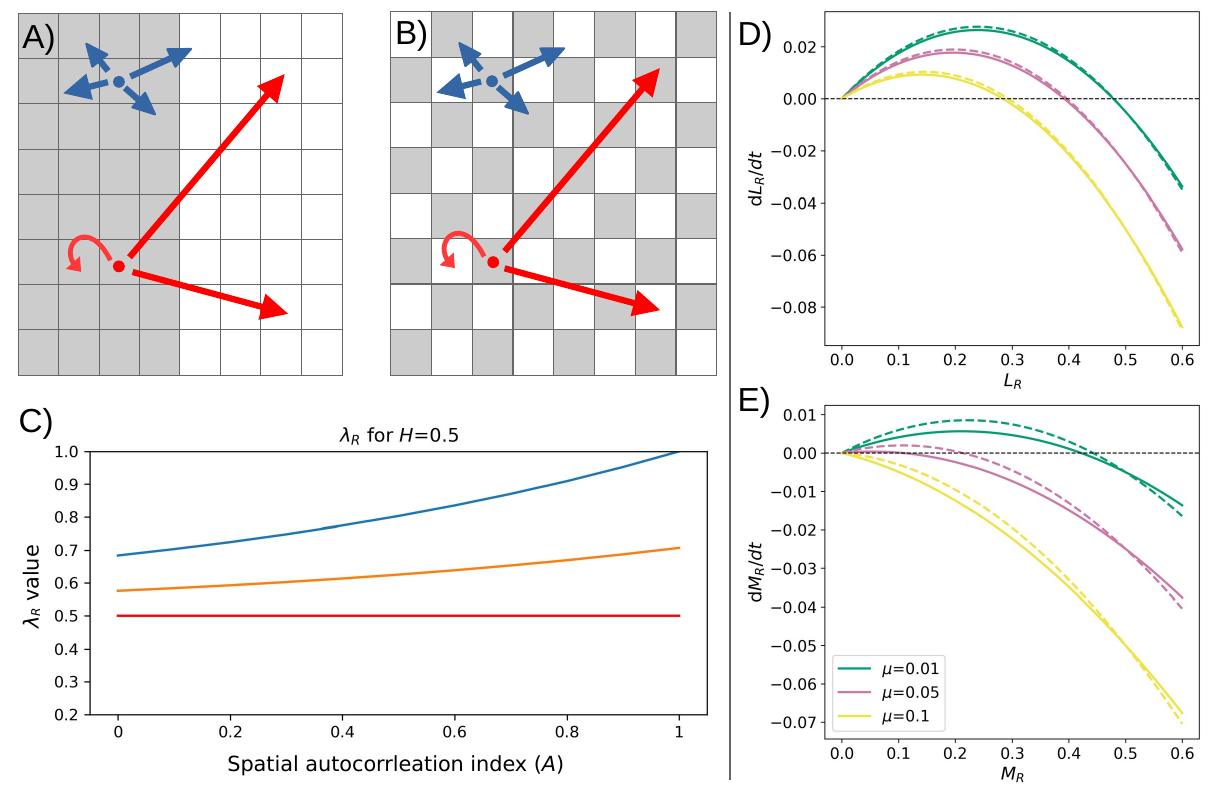}
	\centering
	\caption{\textbf{Diagram representation of the model.} Diagrams represent the distribution of rich (grey) and poor (white) patches for a landscape with $H$ = 0.5 and with either a high ($A$ = 1) (A) or low ($A$ = 0) (B) spatial autocorrelation index. Red circles represent a coloniser species with global dispersal ($d$ = 1), while blue circles are a competitor species with short dispersal ($d$ = 0). In plot (C) we show how the coloniser species (red line) has a $d$ = 1 which corresponds to a $\lambda$ = 0.5 for any spatial autocorrelation index, while the competitor species (blue line; $d$ = 0) has a $\lambda \sim$ 1 in spatially autocorrelated landscapes, but a $\lambda \sim$ 0.68 in non-autocorrelated landscapes (value estimated for a strategy dispersing following a Poisson dispersal kernel centred at a distance of 1 square in a spatial lattice; see main text for details). The orange line is for a species with $d$ = 0.5. Plots (D) and (E) display qualitative results for $L_R$ \eqref{eq:3} when $M_R$ = 0, and $M_R$ \eqref{eq:4} when $L_R$ = 0, respectively. In both cases, we use $H$ = 0.5, and we compare the results with $A$ = 0 (solid line) and $A$ = 1 (dashed line) for three levels of $\mu$ (0.01, 0.05, 0.1).}
	\label{fig:1}
\end{figure}

Habitat fragmentation is associated with a reduction of movements among inhabitable areas, often separated by uninhabitable space, although the influence of this reduction depends on the dispersal strategy. For instance, species with a global (high range) dispersal, may be able to find suitable habitat no matter the distance between inhabitable patches. However, a species with local (short range) dispersal may see its dispersion highly restricted with increasing fragmentation due to the loss of connectivity. Based on these premises, we add a component to our model to incorporate differences in colonisation abilities between competitors and colonisers at different levels of landscape heterogeneity. We developed $\lambda_{Rj}$ (\ref{eq:5}), the rate of landing in habitat $R$ when species $j$ reproduces in a patch of quality $R$. This equation ($\lambda_{Rj}$) is inspired from existing metapopulation and island biogeography models, and depends on three factors: habitat autocorrelation ($A$), proportion of poor patches ($H$) and dispersal distance ($d_j$).

\begin{eqnarray}\label{eq:5}
	\lambda_{Rj} = 1 - H + ((1 - H)^{d_j} - 1 + H) e^{A-1}
\end{eqnarray}

Equation \eqref{eq:5} was developed to approximate the rate of landing in a rich patch for a species following a Poisson dispersal kernel with a mean distance of $X$ squares in a spatial lattice, depending on the habitat autocorrelation and habitat loss indices. Therefore, $\lambda_{Rj}$ can exhibit a variety of behaviours, from linear (for global dispersers) to exponential (for local dispersers). For instance, in a landscape with an equal proportion of rich/poor patches ($H$ = 0.5), this equation returns $\lambda_{Rj}$ = 0.5 for a species with global dispersal ($d_j$ = 1), regardless of the distribution of patches (\cref{fig:1}A, B, C). However, for species $j$ with low $d_j$ (limited dispersal ability), the rate of changing habitat depends heavily on landscape structure (\cref{fig:1}A, B). Thus, $\lambda_{Rj}$ is close to $1 - H$ in a landscape with no spatial autocorrelation ($A = 0$), but tends to 1 when the landscape is highly autocorrelated ($A = 1$) (\cref{fig:1}C). For example, this equation returns $\lambda_{Rj} \sim 1$ in a highly autocorrelated ($A$ = 1) landscape with $H$ = 0.5, but $\lambda_{Rj} \sim 0.68$ in a uncorrelated landscape -- with randomly distributed patches -- for a species with $d_j$ = 0. The latter approximates the true value obtained for a strategy dispersing following a Poisson kernel distribution of 1 square in a lattice with 50\% of randomly distributed rich patches (e.g., chess board-like; $(1 - e^{-1})/2$). Thus, we use equation \eqref{eq:5} as a proxy for the probability of landing in a rich patch for a given species with $d_j$ in a landscape with any $A$ and $H$. In this model, the cost of dispersal for a coloniser species is included in the high probability to land in poor patches, and we ignore additional costs of dispersal.

\subsubsection{Analytical solution for two species:}
We can find an analytical solution for $M_R$ and $L_R$ as long as the two species follow a competitive hierarchy. At equilibrium (i.e., when there is no change in the abundance of either species), we can set equations \eqref{eq:3} and \eqref{eq:4} equal to zero, and find the proportion of patches occupied by a given species $M_R$:

\begin{equation}\label{eq:6}
	M_R = 1 - H - \frac{\mu}{c_{M} \lambda_{RM}}
\end{equation}

as well as for a given species $L_R$, depending on the occupancy of species $M_R$:

\begin{equation}\label{eq:7}
	L_R = 1 -H - M_R -\frac{ c_M M_R \lambda_{RM}+\mu}{c_L \lambda_{RL}}
\end{equation}

With these two equations we can find $M_R$ and $L_R$ for two species, with any combination of $d$ and $c$, in an environment with any $H$, $A$ and $\mu$. On the one hand, we can observe that the occupancy of the competitor species \eqref{eq:6} increases with the amount of suitable habitat, colonisation rate and habitat autocorrelation. Equation \eqref{eq:6} is comparable to the results obtained by \citet{Tilman1994}, ($1 - \frac{m}{c}$), where $m$ equals mortality and $c$ equals colonisation rate. On the other hand, the occupancy of the coloniser species \eqref{eq:7} increases with increasing colonisation ability; the higher the difference in colonising ability between competitor and coloniser, the higher the occupancy of the coloniser species (\cref{fig:s1}A). We use this model to study the conditions supporting coexistence of competitor and coloniser species for different disturbance regimes ($\mu$), spatial autocorrelation levels ($A$), and degrees of habitat loss ($H$). For each scenario, we extract the relative abundance of each species. In cases where there was less than 0.1\% occupancy of a species, we considered it extinct. A basic local stability analysis of \eqref{eq:6} and \eqref{eq:7} reveals that whenever the two species coexist, their coexistence is locally stable (i.e., the population at the equilibrium is robust against a small perturbation in the densities; see Supplementary file 1 for more information). \par
\

\subsection{Extension to $N$ species:}
We can generalize this model to any number of species, as long as species are ordered according to a competitive hierarchy from the strongest to the weakest competitor (sensu \cite{Tilman1994}). We first obtain the proportion of patches occupied by the strongest competitor with \eqref{eq:6}, following which, the patch occupation for any other competitor $M_j$, in decreasing rank of competitive hierarchy, can be determined with:

\begin{equation}\label{eq:10}
	M_j = 1 - H - \sum{M_i} - \frac{N(\sum{M_i c_i \lambda_{Ri}}) + \mu}{c_{j} \lambda_{Rj}}
\end{equation}

where $i$ refers to competitors stronger than $M_j$. Thus, $M_i$, $c_i$ and $\lambda_{Ri}$ respectively refer to the occupied patches, colonisation ability and fraction of offspring remaining in the rich habitat after dispersal, for all stronger competitor species ($N$). We can observe that this equation is equivalent to \eqref{eq:6} when $M_i$ = 0, that is for patch occupancy of the strongest competitor. In this case, we use the Shannon effective diversity index to quantify biodiversity (coexistence of multiple species). As for the two-species case, we consider a species extinct if it has less than 0.1\% occupancy.

\section{Results}
First, we investigate the coexistence between two species with clear differences in colonisation and competition abilities to demonstrate the behaviour of the model for a classical CCTO scenario. We refer to these two species as the coloniser (high dispersal distance and colonisation rate, but with low competitive ability) and competitor species (low dispersal distance and colonisation rate, but high competitive ability). For simplicity, following the premises of CCTO, we use $c = d$, meaning that species that disperse often also disperse farther. We also explore the influence of relaxing this assumption and considering colonisation rate (or fecundity) and dispersal distance separately in the supplementary material (\cref{fig:s1,fig:s2}), given that previous studies of plants showed that coloniser species with fast-growing exploitative strategies could benefit from short-dispersal \citep{Bolker1999, Bolker2003a, Snyder2003}. Second, we investigate the effects of considering multiple species, and assess the biodiversity patterns across habitat loss and disturbance gradients. For this purpose, we run the model with increasing numbers of species (5, 10, 50) using values of $c$ and $d$ equally spaced between 0.1 and 1, forming a competitive hierarchy.
    
\subsection{Results with two species}
We report the results obtained for two species with $c_i = d_i$, for three levels of disturbance ($\mu$ = 0.01, 0.05 and 0.1), and analyse the impact that varying habitat loss ($H$) and spatial autocorrelation index ($A$) have on the species coexistence. We also present the main differences between this case and one considering species with differing in colonisation rate but with both species restricted to local dispersal ($d_L$ = $d_M$ = 0.2; see \cref{fig:s1,,fig:s2}).\par
\

\begin{figure}[H]%
	\includegraphics[width=\textwidth]{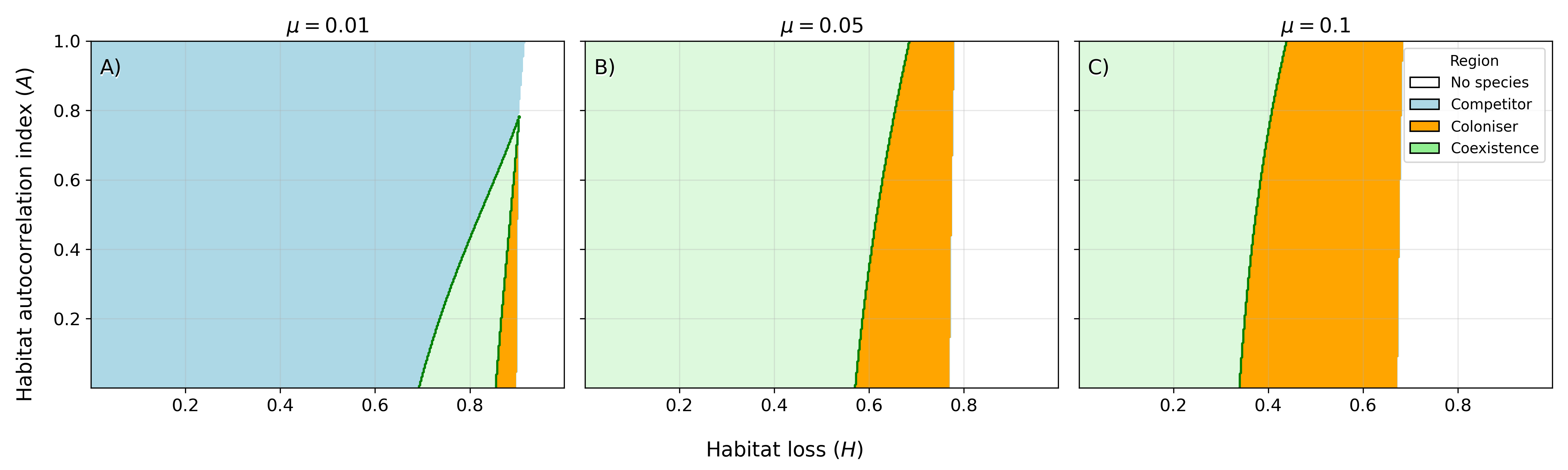}
	\caption{\textbf{Coexistence patterns for two species.} Effects of spatial autocorrelation index ($A$), and proportion of habitat loss ($H$) on coexistence between a competitor ($c_M$ = $d_M$ = 0.2) and a coloniser ($c_L$ = $d_L$ = 0.9) with a competitive hierarchy. The mortality rate due to disturbances ($\mu$) is indicated on top of each plot. White cells indicate that no species survived. Results were computed using \eqref{eq:6} and \eqref{eq:7}, and the analytical result of the stability analysis was consistent with the coexistence area.}
	\label{fig:2}
\end{figure}

\subsubsection{Habitat loss ($H$):}
Habitat loss has a clear effect on coexistence of species, although its effect depends on the interaction with disturbance (\cref{fig:2}). Increasing habitat loss generally benefits colonisers, and it is at intermediate levels of disturbance ($\mu$ = 0.05) that both species coexist over the broadest parameter space. With increasing disturbance, coexistence is only possible with low levels of habitat loss.\par 

\subsubsection{Disturbance ($\mu$):}
The resilience of coloniser species increases with $\mu$, with $\mu$ = 0.05 being the scenario that supports coexistence over the broadest parameter space (\cref{fig:2}). At low levels of disturbance, the competitor species always excludes the coloniser species, but colonisers have higher relative success in scenarios with high disturbance. However, it is worth noting that in scenarios with high habitat loss, even intermediate levels of disturbance lead to general extinction (\cref{fig:2}B).\par 

\subsubsection{Spatial autocorrelation ($A$):}
The effects of spatial autocorrelation are dependent on the level of disturbance and habitat loss. In general, increasing autocorrelation selects for competitor species and plays a minor role in determining species coexistence. However, it can play a critical role in specific landscape compositions: low disturbance level and high habitat loss (\cref{fig:2}A), or intermediate/high levels of disturbance ($\mu$ = 0.05 or 0.1) and a balanced proportion of rich and poor patches (\cref{fig:2}B, C).\par

\subsubsection{Two species with local dispersal:}
The results obtained with competitor and coloniser species with equal and limited dispersal ability were broadly comparable with the case described above. A notable contrast was that local dispersal benefitted colonisers in landscapes with high spatial autocorrelation and high habitat loss, compared to colonisers with almost global dispersal (\cref{fig:s2}). In terms of coexistence, local dispersal had a limited effect under intermediate-high disturbance, but enhanced coexistence in habitats with high habitat loss and $\mu$ = 0.01 (\cref{fig:s2}). See \cref{fig:s1} for more information on the asymmetric interactive effects of the model when uncoupling the colonisation rate (fecundity) and the dispersal distance.\par

\subsection{Results with $N$ species:}

Results for five species with different colonising abilities are broadly comparable to those of two species: with low disturbance, coexistence is limited to landscapes with high habitat loss, while an intermediate disturbance rate supports a larger range of coexistence conditions, and coloniser species benefit from disturbed and fragmented landscapes (\cref{fig:s3}). To better illustrate the relative effect of spatial autocorrelation on biodiversity across the habitat loss gradient, we computed the variance of the entropy index induced by changes in $A$, and this was done for each value of $H$. If spatial autocorrelation has no effect on biodiversity, the variance induced is zero. In the case of five or 10 species, spatial autocorrelation plays a role only in landscapes with a high proportion of habitat loss when disturbance levels are low -- similar to the two-species case. However, its effects expand over a larger habitat loss gradient as the number of species increases (50 species), particularly in landscapes with intermediate/high level of disturbance (\cref{fig:3}). Thus, while spatial autocorrelation plays a minor role in shaping biodiversity compared to habitat loss or disturbance, it becomes more influential in determining coexistence of highly diverse communities under intermediate to high disturbance conditions ($\mu$ = 0.01 or 0.05).\par 

\begin{figure}[H]%
	\includegraphics[width=\textwidth]{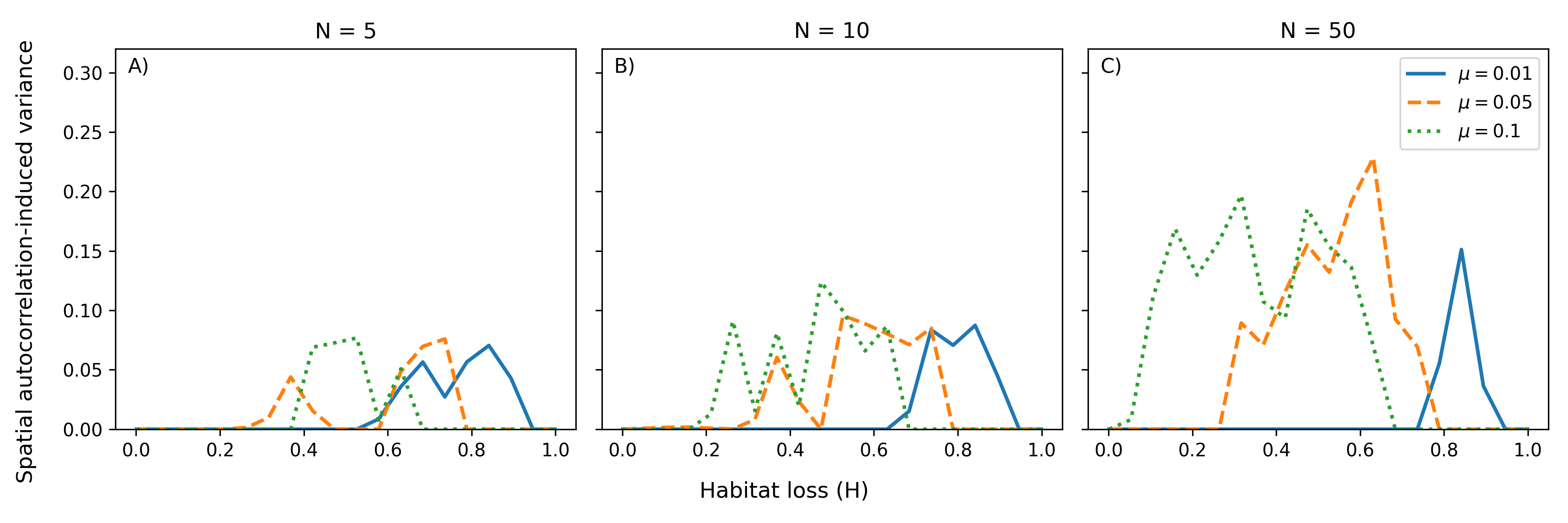}
	\caption{\textbf{Influence of spatial autocorrelation on biodiversity.} Magnitude of the effects of spatial autocorrelation ($A$) depending on the proportion of habitat loss ($H$) and disturbance rate ($\mu$; indicated by different types of line). For the spatial-autocorrelation-induced variance (Y-axis) we gathered the Shannon effective diversity index obtained for all levels of $A$ (from 0 to 1) and a single value of $H$, and computed the variance of the indexes for each level of $H$. When varying $A$ has no effect on biodiversity, the Y-axis is zero, but it increases in value with increasing variance. We ran the model for A) five, B) 10 and C) 50 species, as indicated on the top of each panel. In all cases, species were given a matching competitive and dispersal ability ($d_i$ = $c_i$), and these values were equally spaced between 0.1 and 1, forming a competitive hierarchy.}
	\label{fig:3}
\end{figure}

The results for five (\cref{fig:s3}), and 10 (\cref{fig:4}), species reveal oscillatory patterns of biodiversity in response to variation in the degree of habitat loss. In general, decreasing values of the spatial autocorrelation index ($A$) lead to biodiversity peaking at lower proportions of habitat loss (\cref{fig:4}). In addition, while habitat loss generally provides a benefit to colonisers, the community of coexisting species does not necessarily appear in a sequential order of colonisation ability. This means that over a gradient of increasing habitat loss, species with high colonisation ability ($c$) may increase in abundance at lower habitat loss levels than a coloniser with a lower $c$ value (\cref{fig:4} and \cref{fig:s3}). For instance, in \cref{fig:4}G, we observe that a coloniser with a $c_i = 0.7$ can coexist with a competitor ($c_i$ = 0.2) at $H \sim 0.22$, whereas species with colonisation abilities of 0.3 - 0.6 only coexist with this competitor when $H > 0.23$. Thus, increasing habitat loss does not directly imply a benefit to the species with the highest dispersal range, but can benefit intermediate species in some cases (\cref{fig:s3}). While we are interested in biodiversity, it is also important to understand whether the environment selects for coloniser or competitor strategies. We observe that the species community largely differs among peaks of similar diversity, such as species with high dispersal range ($c_i \ge 0.6$) are present only when the proportion of habitat loss and/or disturbance level is high (\cref{fig:4}).\par 
\

\begin{figure}[!h]%
	\includegraphics[width=\textwidth]{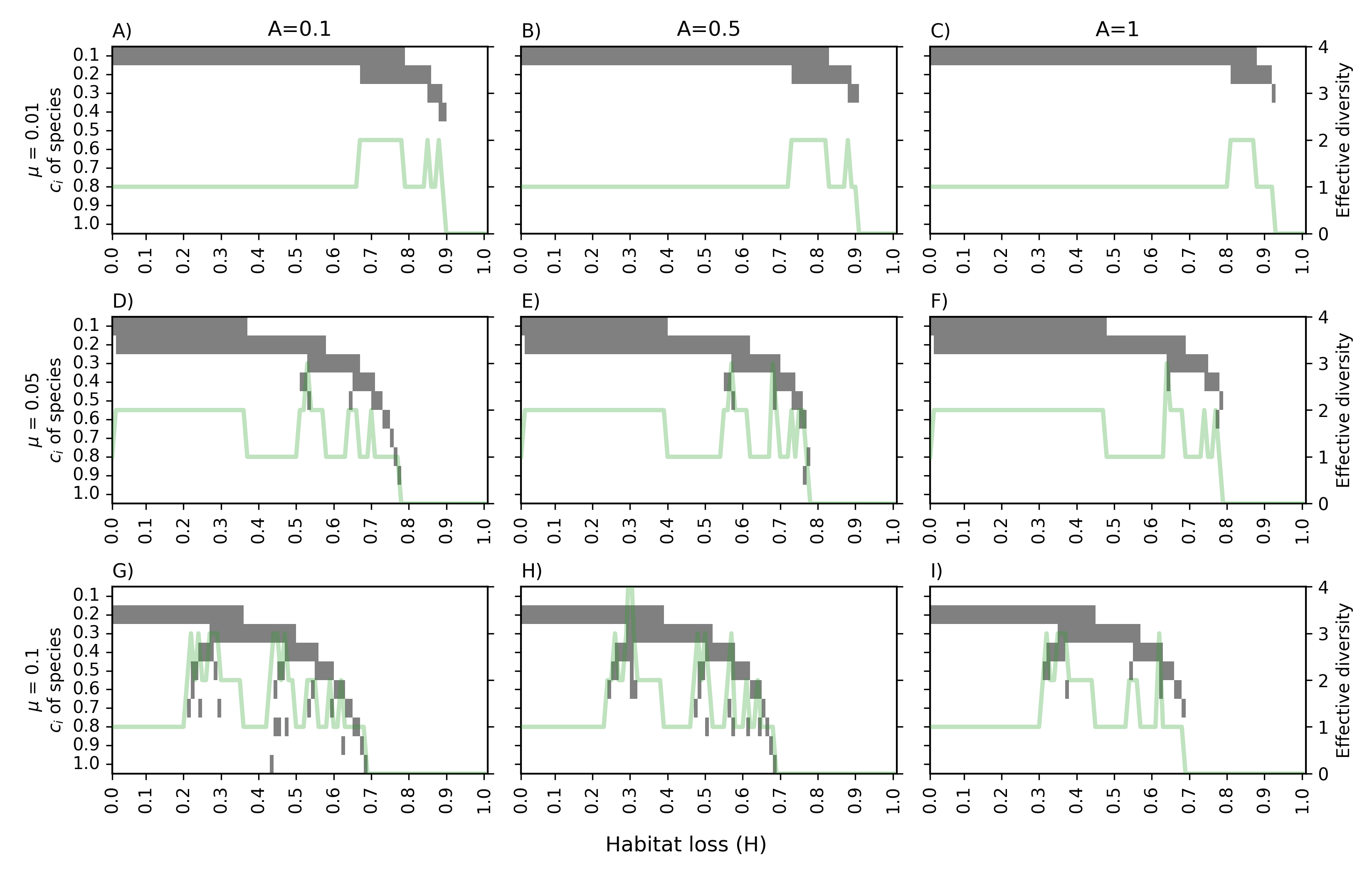}
	\caption{\textbf{Co-occurrence of species and effective diversity for different environmental conditions.} Results for the analytical model with 10 species, assuming a competitive hierarchy, with the competing/colonising abilities for each species $i$ ($c_i$ = $d_i$) equally spaced between 0 and 1 (left Y-axis). The grey cells indicate the presence of a species over a gradient of habitat loss, while white cells indicate its absence. The right Y-axis represents the effective diversity, and is plotted in green. The X-axis is the proportion of habitat loss ($H$). The disturbance level ($\mu$) is indicated on the left, while the spatial autocorrelation ($A$) is indicated on top.}
	\label{fig:4}
\end{figure}

In the case of 10 species, we also investigated the biodiversity patterns across a disturbance gradient. A variety of biodiversity-disturbance patterns appear for different combinations of habitat loss and spatial autocorrelation index (\cref{fig:5}). While some parameter combinations provide support for the IDH, we also observe multi-peak patterns when there is some degree of habitat loss (e.g., $H$ = 0.4).\par 
\

\begin{figure}[!h]
	\includegraphics[width=\textwidth]{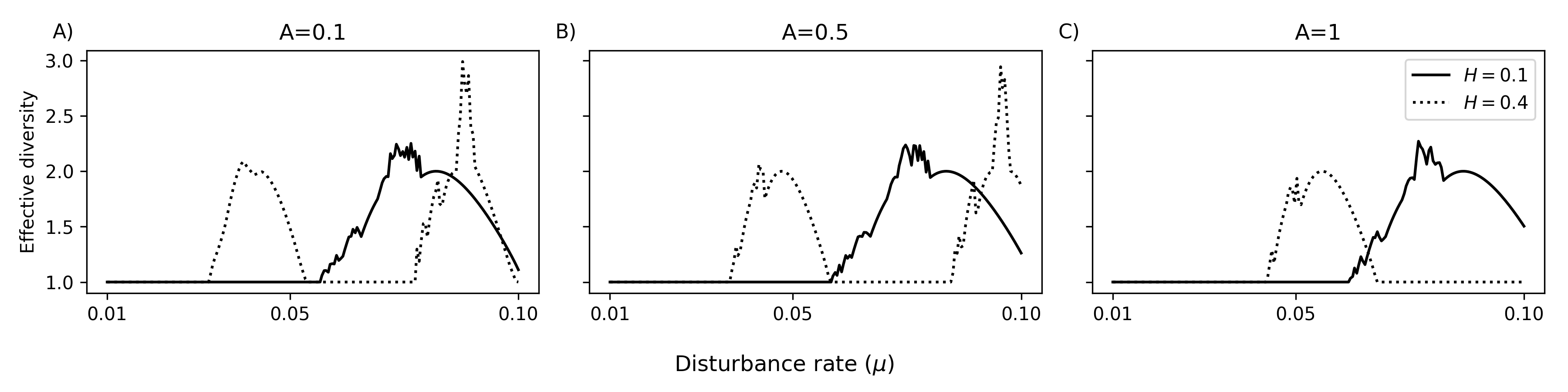}
	\caption{\textbf{Diversity-disturbance patterns for 10 species}. Figures show the effective diversity (Y-axis) depending on the mortality rate due to disturbance level (X-axis). We show diversity patterns for three levels of spatial autocorrelation (indicated on top of each panel, A-C). In all cases we show the results of the model for scenarios with low ($H$ = 0.1; continuous line) and intermediate ($H$ = 0.4; dotted line) levels of habitat loss.}
	\label{fig:5}
\end{figure}

In the case of 50 species, the addition of species with intermediate competitive/colonising abilities increase the frequency of biodiversity oscillations in response to variation in habitat loss, relative to 10 or fewer species (\cref{fig:6}). This expands the range of habitat that supports diversity, producing more biodiversity peaks along the environmental gradient. However, the overall trend in biodiversity is a decreasing one when the disturbance rate is high, but increasing when the disturbance rate is low, regardless of the spatial autocorrelation index (\cref{fig:6}). Such differences in oscillatory patterns suggest a complex interaction between the influence of environmental factors and the CCTO. As an example, landscapes with low spatial autocorrelation index ($A$ = 0.1) that were dominated by a single competitor species when considering 10 species (\cref{fig:4}G), can support coexistence of multiple species when considering 50 species (\cref{fig:6}A). These oscillation patterns have important implications for ecological studies trying to provide empirical evidence for theoretical models. Researchers conducting field studies tend to sample the environment in a discrete gradient (e.g., N plots in three differently disturbed areas). When taking three evenly separated biodiversity measures along a gradient of disturbance (0.01, 0.05 and 0.1), our model can generate all kinds of biodiversity patterns depending on fragmentation levels (\cref{fig:s4}). This is due to the oscillations of the model (\cref{fig:s4}), which are shifted depending on fragmentation (habitat loss and spatial autocorrelation). It should be pointed out that we consider species with $c_j = d_j$ here, and considering alternative species with $c_j \ne d_j$ could result in different coexistence dynamics, although this is beyond the scope of this study.\par

\begin{figure}[H]
	\includegraphics[width=\textwidth]{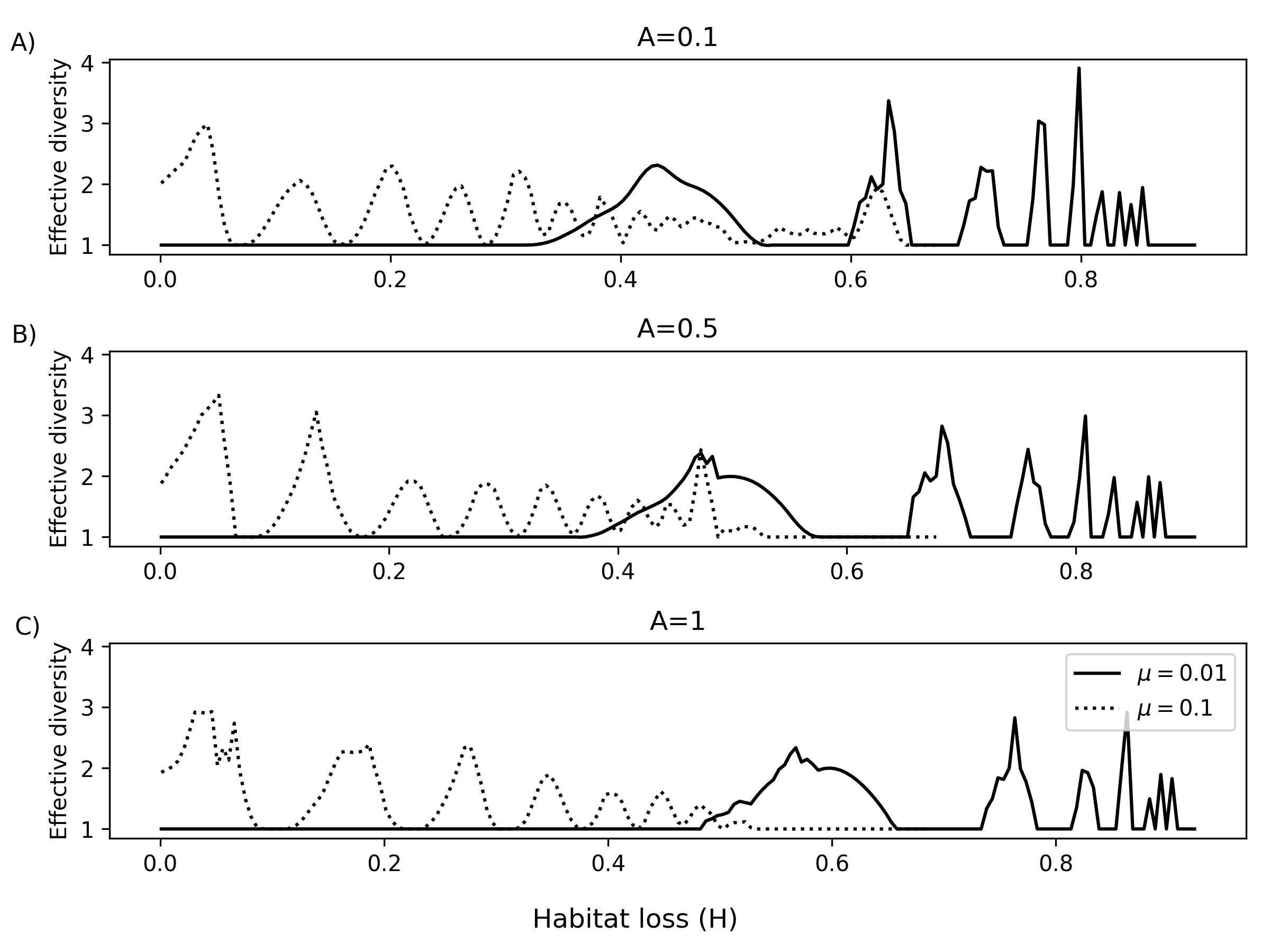}
	\caption{\textbf{Effective diversity curves.} Results for the analytical model with 50 species, with their competing/colonising abilities ($c_i$ = $d_i$) equally spaced between 0 and 1 assuming a competitive hierarchy. The Y-axis indicates the effective diversity index while the X-axis is the \% of habitat loss ($H$). Each line type corresponds to a different mortality rate ($\mu$) due to disturbance. The spatial autocorrelation ($A$) is indicated on top of each plot.}
	\label{fig:6}
\end{figure}

\section{Discussion}
Theoretical studies can make a valuable contribution to species conservation through predicting how human disturbances and climate change may affect the distribution and survival of species \citep{Kubisch2013, Travis2012, Bonte2012}, but face challenges in modelling the complexity of natural systems. Here, we present a simple new method to integrate habitat loss and spatial autocorrelation with the competition-colonisation trade-off, one of the most influential models of spatial niche partitioning \citep{Yu2001}, to test general relationships between biodiversity (coexistence of species) and key environmental drivers: habitat fragmentation and disturbance. In validation of the approach, our findings are generally consistent with previous theoretical research on the dynamics of competitor/coloniser species interactions and the influence of single environmental components on species coexistence. First, we observed that coloniser strategies dominate in scenarios with high disturbance level, and short dispersal can increase biodiversity when the disturbance level is low \citep{DosSantos2011, Bolker1999, Bolker2003a, Snyder2003}. Second, we observe that spatial autocorrelation plays an important role in supporting coexistence at intermediate-high levels of habitat loss ($H \ge$ 0.4), allowing coloniser species to occupy zones that are difficult to access by competitor species. This result is in agreement with previous studies showing that while spatial autocorrelation can act as a stabilising mechanism and promote coexistence via niche differentiation \citep{Chesson2000,Liao2022, Levine2017}, its effect is weak if there is no habitat loss \citep{King2002, Planas-Sitja2023, Cronin2016, Mathias2001b}. Third, we find general support for the intermediate disturbance hypothesis (IDH) as a mechanism promoting species coexistence \citep{Chesson2000,Liao2022, Gross2008, Cadotte2007, DosSantos2011}, but our model also generates alternative biodiversity patterns (e.g., multi-peak) across the habitat loss gradient, including oscillatory dynamics previously described in the literature \citep{Liao2022, Li2020, Zhang2023}. \par 
\

Notably, our results highlight the importance of interactive effects among different factors influencing species coexistence. Previous work suggested that communities subjected to a CCTO (with strict competitive hierarchy) were expected to display uni-modal biodiversity-disturbance curves \citep{DosSantos2011}. While our results broadly support this IDH prediction, we find that different biodiversity patterns emerge at fixed values of habitat autocorrelation or habitat loss when varying disturbance levels (e.g., monotonically increasing, bimodal). These different patterns become even more pronounced when considering a higher number of species. It is possible that, in the case of few species, some of them possess a near-optimal strategy for a specific environment (combination of disturbance, habitat loss and spatial autocorrelation index), and thus dominate in such a scenario. With more species, however, the parameter space explored expands, and a subset of species (competitors, colonisers or both) can display strategies with similar fitness and thus coexist. These interactive effects between the CCTO, fragmentation level and disturbance, could thus be important to help explain the alternative biodiversity-disturbance relationships that have been reported in nature \citep{Lenz2004, Cadotte2007, Hall2012}.\par 
\

Recent theoretical studies showed that oscillatory biodiversity patterns in response to variation in habitat loss or disturbance could arise naturally with the introduction of several species under a CCTO hierarchy, where the rise, or decline, of the best competitor precluded, or facilitated, the coexistence of a specific subset of species \citep{Zhang2023, Liao2022, Li2020}. This multi-modal pattern was enhanced by the introduction of intransitive competition (such as
in the rock-paper-scissors game), as opposed to hierarchical competition, via stabilising mechanisms \citep{Li2020, Zhang2023}. We found similar oscillatory biodiversity patterns across the habitat loss gradient, but these oscillations were only visible across the disturbance gradient when considering a high number of species (50 species). Because it is not feasible to sample a continuous environment in the field as theoretical models do, these oscillations provide a potential explanation for how discrete sampling can yield different biodiversity-disturbance curves depending on the number of species and habitat configuration. For instance, at a fixed level of habitat autocorrelation or habitat loss, our model shows that sampling three levels of disturbance intensity can produce increasing, decreasing or uni-modal biodiversity patterns across the gradient \cref{fig:s4}. While our approach shares elements with prior work (notably \citet{Zhang2023}), the introduction of the $\lambda$ factor in our model offers two key advantages: it enables dispersal across habitat fragments, and ensures that the effect of connectivity loss scales with species dispersal capability. This addition may explain why species coexistence does not necessarily follow a sequential order of increasing colonisation ability with increasing disturbance intensity or habitat loss proportion, unlike previous studies \citep{Zhang2023, Liao2022}. It is important to note that, even in the presence of oscillatory responses, the overall diversity tends to decrease with habitat loss when the disturbance level is high, but increases when disturbance is low. This result is consistent with previous theoretical work on CCTO communities -- although not explicitly focused on fragmentation -- which showed that diversity decreases with productivity (comparable to our proportion of habitat) when disturbance is low ($\mu \sim 0.01$) but increases when disturbance is high ($\mu \ge 0.1$) \citep{Kondoh2001}.\par 
\

Studies summarizing the effects of habitat fragmentation independent of habitat amount, equivalent to our spatial autocorrelation index, found that fragmentation had positive effects on biodiversity in most cases (76\% of significant responses found in the literature) \citep{Fahrig2017,Riva2023,Riva2022}. These results suggested that fragmented landscapes -- similar area but in small patches -- should receive the same conservation value as an equivalent area within a large patch. Our theoretical approach supports this idea because, in many cases, decreasing spatial autocorrelation, for the same amount of habitat, leads to higher biodiversity (\cref{fig:5} or \cref{fig:6} for $H$ = 0.4). We also find that the effects of spatial autocorrelation are highly dependent on habitat available and disturbance level, such that it is less influential where disturbance is low and inhabitable space is abundant. Secondly, while the effects of spatial autocorrelation are weak relative to habitat loss, spatial autocorrelation can significantly shape the consequences of habitat loss, especially for intermediate values of habitat loss (\cref{fig:3,,fig:4}). With this in mind, we can extrapolate the findings of our model to real-world contexts. We predict unimodal biodiversity-disturbance patterns in a relatively small green area (e.g., small forest parks, urban forests) ($H \sim 0.4$) with high connectivity ($A \sim 1$), but multimodal patterns if the same area is fragmented into smaller patches (e.g., small parks, urban gardens) ($A \sim 0.1$). In addition, this study suggests that exceeding a 70\% threshold of habitat loss in low disturbed areas could lead to the survival of coloniser-only species profiles (if dispersal distance allows moving to scattered habitable areas), or in highly disturbed areas (e.g., use of pesticides, tourism, gardening) could lead to extinction (\cref{fig:4,,fig:6}).\par
\

The inherent flexibility of our model offers several benefits when investigating the effects of biological traits and environmental factors on species coexistence. Firstly, we can uncouple the colonisation rate (fecundity) from the dispersal distance for any number of species, allowing us to investigate the effects of shifting the dispersal advantage of the coloniser species (from global to local dispersal). In this context, despite the differences between our model and previous models using neighbouring dynamics or moment equations \citep{Bolker1999, Bolker2003a, Snyder2003}, our model qualitatively reproduces their finding that a high colonisation advantage can support coexistence with a competitive species that has the same local dispersal -- but only in landscapes with high spatial autocorrelation and high habitat loss (\cref{fig:s1,fig:s2}). A possible explanation is that the high fecundity allows the coloniser species to exploit empty patches quickly before competitor occupation, in addition to the advantage of a decreased mortality associated with local dispersal (i.e., lower chances to land in poor patches) in spatially autocorrelated scenarios \citep{Olivieri1995, Travis1999a, Bonte2010, Cote2017}. Disturbance, on the other hand, generates new empty patches each generation that confer special advantage to coloniser species (when considering a CCTO), and thus acts more like an equalizing rather than a stabilising mechanism, as disturbance slows competitive exclusion \citep{Chesson2000, Liao2022, Levine2017, Fox2013}. However, when there is no dispersal advantage ($\lambda_{RM} = \lambda_{RL}$), disturbance can act as a stabilising mechanism, limiting both species and promoting coexistence over a broader parameter space. Secondly, our approach allows the potential incorporation of additional factors that can influence coexistence, such as different mortality rate for each species, limited amount of resources in rich or poor patches, or even more than two types of patches, which may introduce additional levels of non-linearity and add more realism to the model, though we leave these possible extensions to future studies. Finally, while we limit our investigation here to the general conditions enhancing coexistence, this system of equations could for example be calibrated with empirical data for specific target species (e.g., manipulating the colonisation rate, dispersal distance, competitive ability, mortality), and use real environmental parameters to predict local biodiversity hotspots or identify threatened species in a given region. In summary, although this model is intentionally simplistic to accommodate a broad range of species, it yields qualitative results consistent with previous literature while generating new insights into how components of habitat fragmentation can interactively impact biodiversity. This approach introduces a new method to integrate spatial heterogeneity into a niche partitioning model, which may have heuristic extensions in combination with other existing models, such as the lottery model, diffusion model or spatial moment equations. \par

\section{Acknowledgements}
We thank Jean-Louis Deneubourg and Nicolas Loeuille for their valuable comments in the early stages of the manuscript.

\section{Funding}
This study was funded by the Japan Society for the Promotion of Science: one  KAKENHI to ALC and IPS (18F18806), one KAKENHI to IPS (23K14229), and two KAKENHI to RI (24H01528, 24H02291).

\bibliography{./TheoreticalEco2025.bib} 

\pagebreak
\nolinenumbers

\begin{center}
	\textbf{\LARGE Supplementary Material: Disentangling the interactive effects of anthropogenic disturbances on biodiversity}
	\\
	
	\author{Isaac Planas-Sitjà*$^{1, 2}$, Ryosuke Iritani $^1$ \& Adam L. Cronin$^2$}
\end{center}

{\parindent0pt
	
	* correspondent author
	\\
	$^1$ RIKEN Wako Campus, Division of Fundamental Mathematical Science, 2-1 Hirosawa, Wako, Saitama 351-0198 Japan
	
	$^2$ Tokyo Metropolitan University, Department of Biological Sciences, 1-1 Minami-Osawa, Tokyo, Japan 
	}

\renewcommand{\thefigure}{S\arabic{figure}}
\crefname{figure}{figure}{figures}
\setcounter{figure}{0}

\begin{figure}[H]%
	\captionsetup[subfigure]{labelformat=empty}
	\begin{subfigure}{1\textwidth}
		\caption{1) Long dispersal}
		\includegraphics[width=\textwidth]{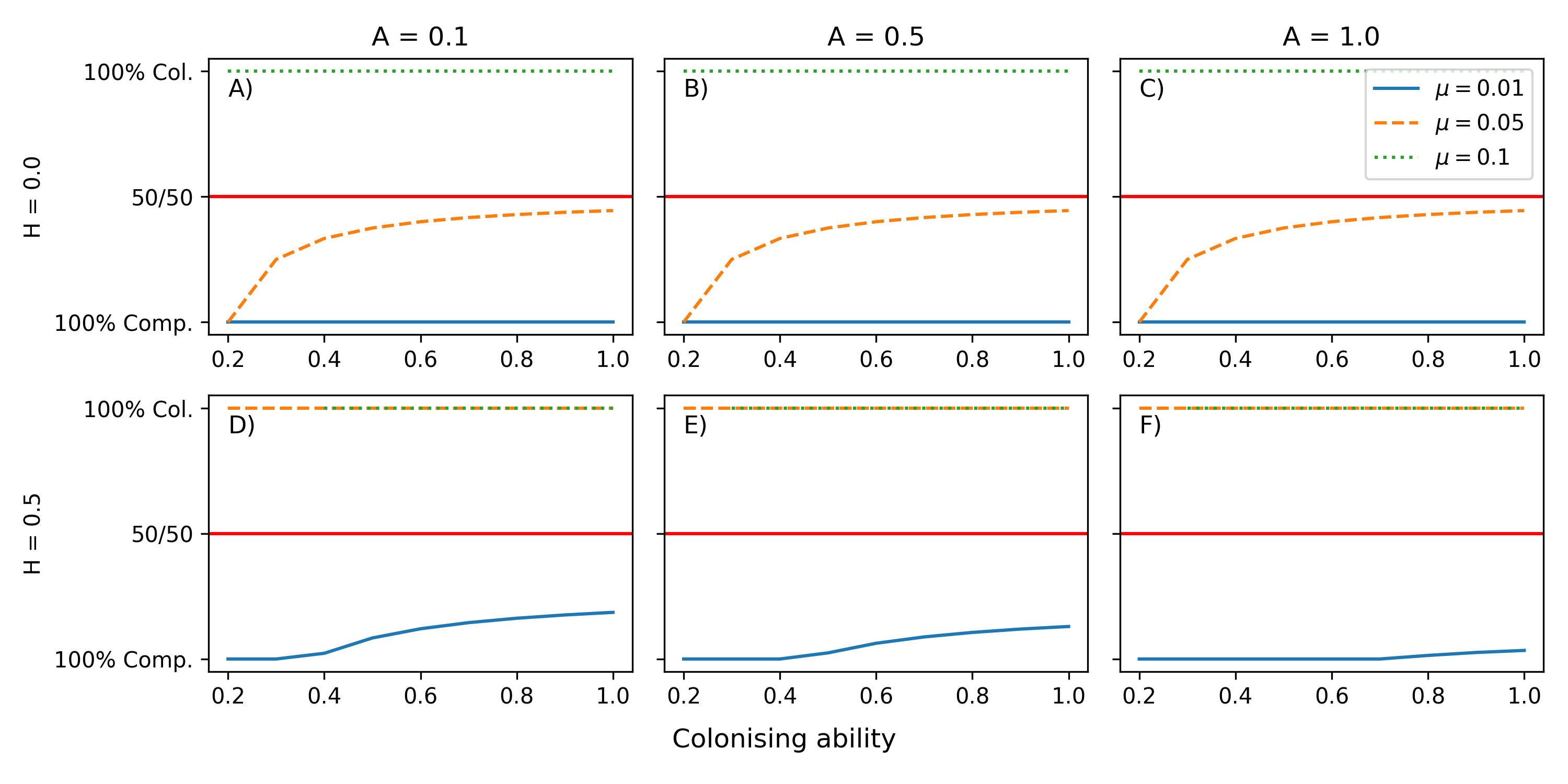}
		\label{fig:s1a}
	\end{subfigure}
	\begin{subfigure}{1\textwidth}
		\caption{2) Short dispersal}
		\includegraphics[width=\textwidth]{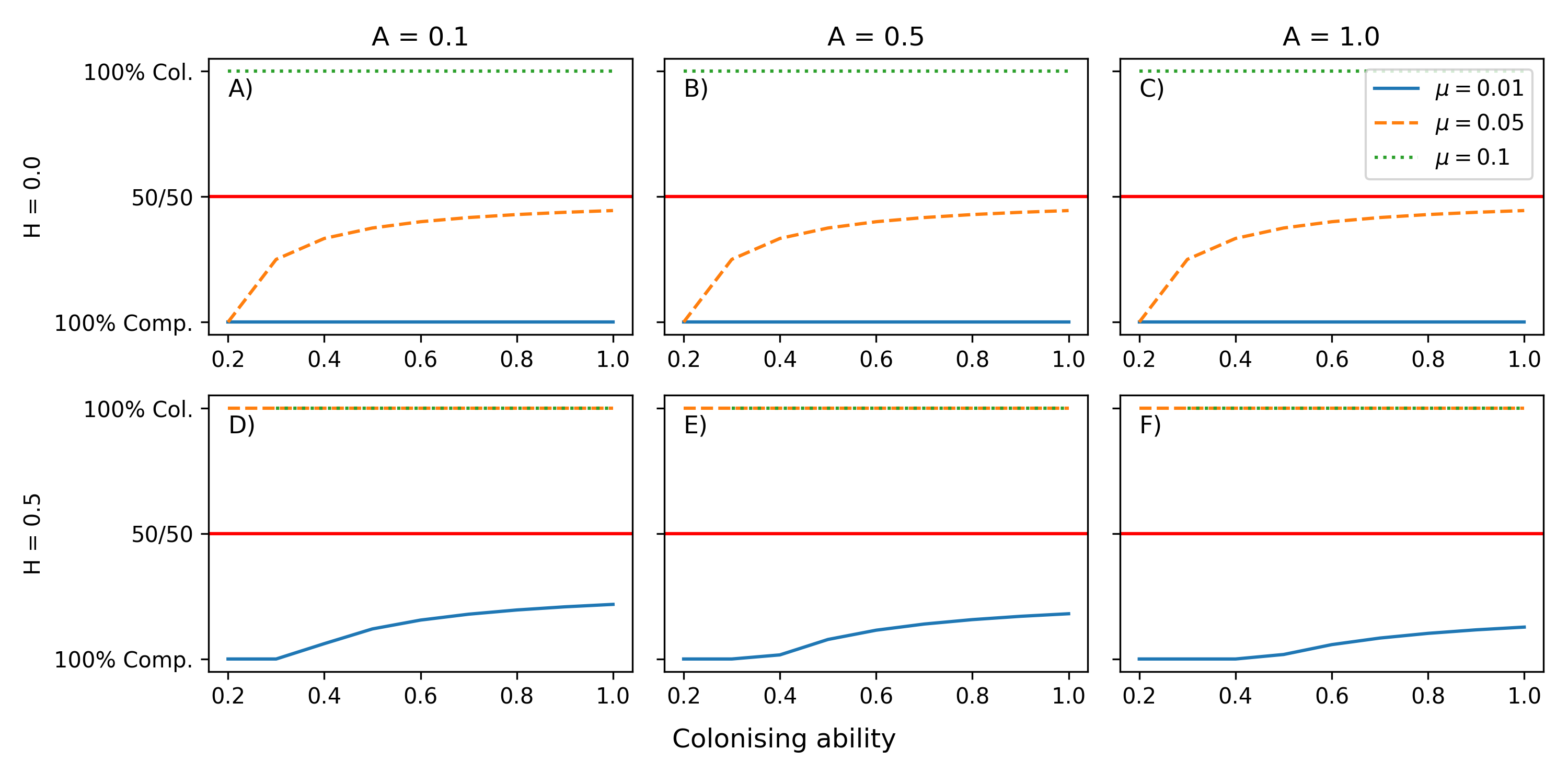}
		\label{fig:s1b}
	\end{subfigure}
	\caption{\textbf{Comparison of results for pairs of species considering long or short dispersal.} Effects of increasing the asymmetry between a fix competitor species ($c_M = 0.1, d_M = 0.1$) versus a coloniser with increasing colonising ability (X-axis). In the case of long dispersal, colonisers have matched fecundity and dispersal ability (($c_L = d_L$), while in the scenarios with short dispersal, dispersal is kept constant ($d_L$ = 0.1) but have increasing colonising ability ($c_L$; X-axis). Y-axis indicates whether competitor and colonising co-existed (red line shows the 50\% co-existing threshold), or whether only the coloniser or competitor were present (100\% Col. or 100\% Comp. respectively). We explore coexistence for three fixed levels of fragmentation (the spatial autocorrelation is indicated on top), and two fixed levels of habitat loss: $H$ = 0 (A - C) or $H$ = 0.5 (D - F). Each line style corresponds to a different mortality rate ($\mu$) due to disturbance level.}
	\label{fig:s1}
\end{figure}

\begin{figure}[H]%
	\captionsetup[subfigure]{labelformat=empty}
	\begin{subfigure}{1\textwidth}
		\caption{1) Long dispersal}
		\includegraphics[width=\textwidth]{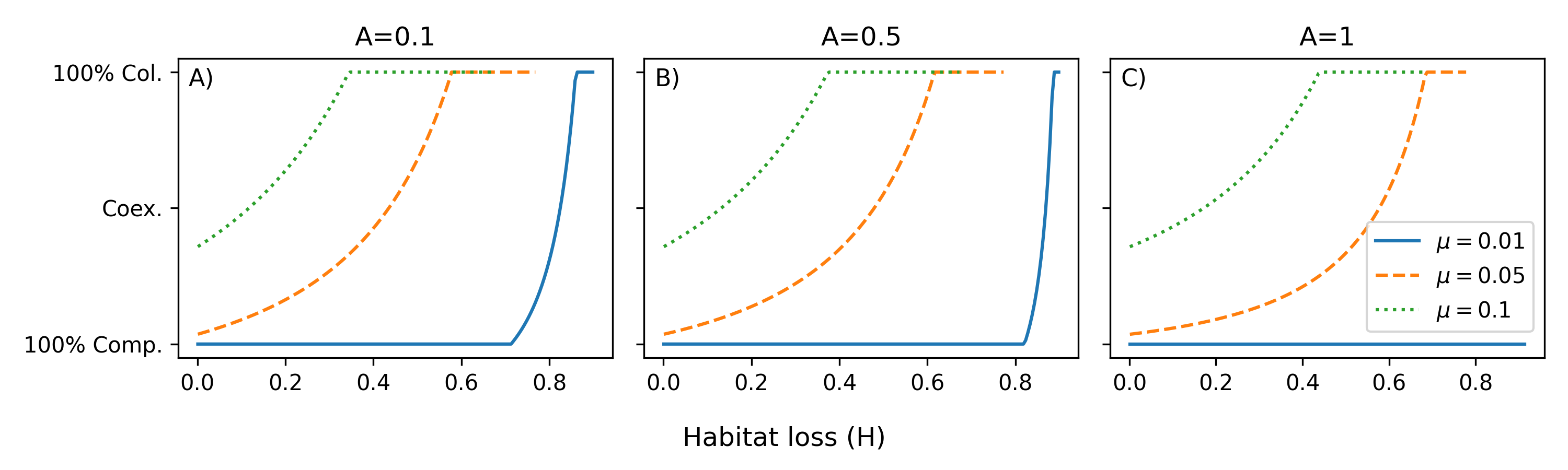}
		\label{fig:s2a}
	\end{subfigure}
	\begin{subfigure}{1\textwidth}
		\caption{2) Short dispersal}
		\includegraphics[width=\textwidth]{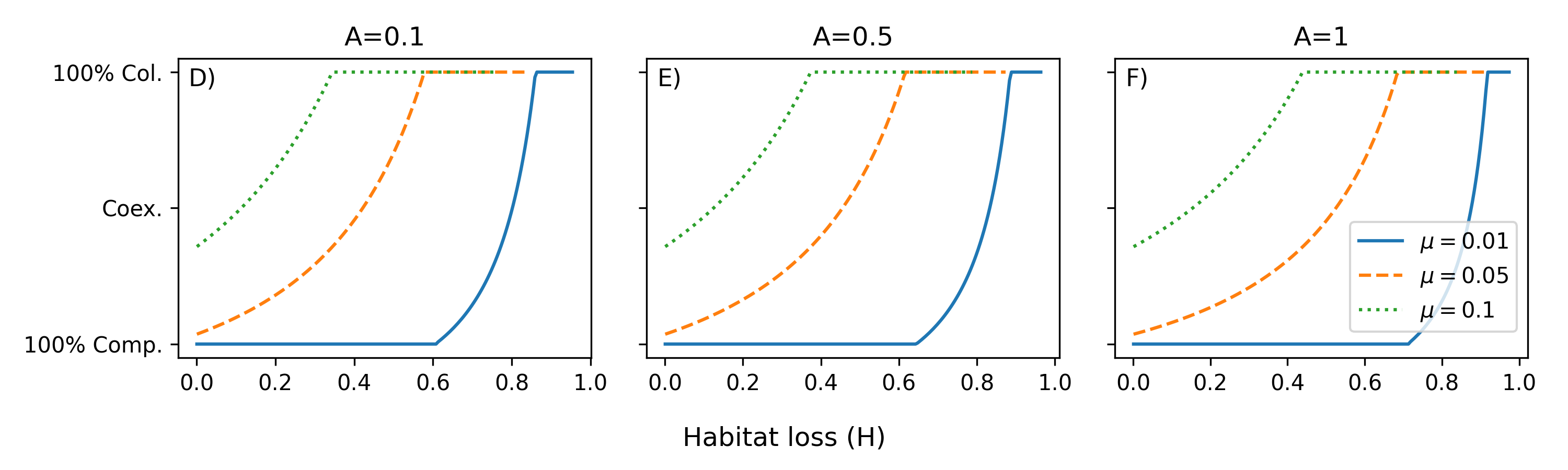}
		\label{fig:s2b}
	\end{subfigure}
	\caption{\textbf{Comparison of results for two species considering long or short dispersal for coloniser species.} Effects of habitat loss ($H$) on coexistence for three fixed levels of fragmentation (the spatial autocorrelation is indicated on top of each figure) for two species with CCTO considering long ($c_L = d_L = 0.9$; A-C) or short ($c_L = 0.9, d_L = 0.1$; D-F) dispersal for the coloniser species, while the competitor species maintains same values ($c_M = 0.1, d_M = 0.1$). Each line style corresponds to a different mortality rate ($\mu$) due to disturbance level.}
	\label{fig:s2}
\end{figure}

Results for pairs of species with increasing differences in colonisation rate (fecundity) and dispersal distance (case 1), or restricting colonisers to local dispersal (case 2), are consistent with the findings in the main text (\cref{fig:s1,,fig:s2}). Results a) provide general support of the IDH when there is no  habitat loss, indicate that b) high mortality selects for the species with largest dispersal range (highest $d$ value), c) increasing dispersal asymmetry between species improves coexistence, and d) considering high colonisation rate linked to local dispersal increases coexistence in scenarios with high habitat loss and low disturbance level.

\begin{figure}[H]%
	\includegraphics[width=\textwidth]{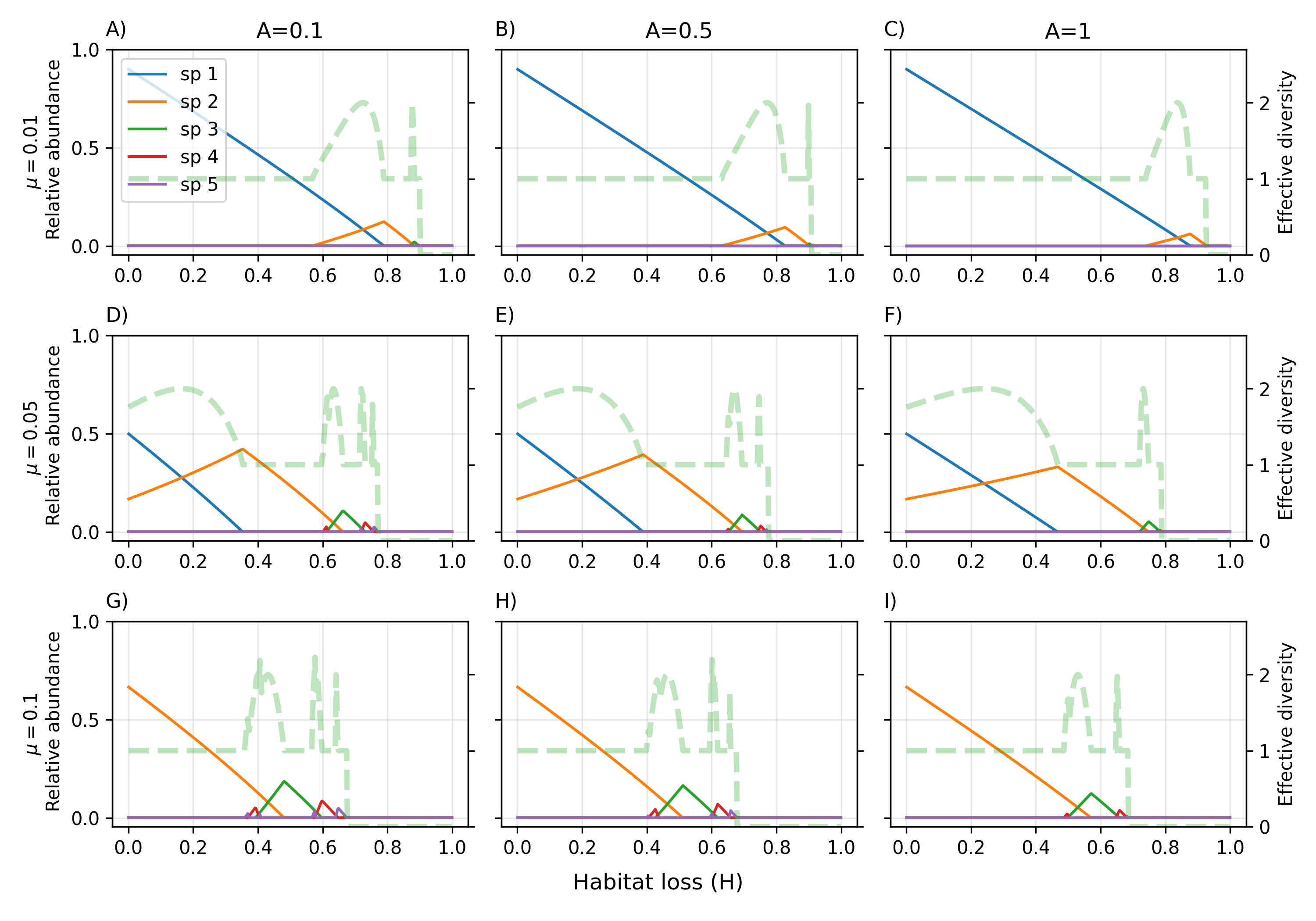}
	\caption{\textbf{Oscillatory diversity patterns for five species.} We explore the coexistence of five species with their competing/colonising abilities ($c_i$ = $d_i$) equally spaced between 0 and 1 assuming a competitive hierarchy, along a gradient of habitat loss ($H$; X-axis). We tested three levels of disturbance level ($\mu$), shown on the left, and three levels of spatial autocorrelation ($A$), indicated on top. The left Y-axis is the relative abundance of each species (proportion of species present in the landscape), and each species is depicted with a different line colour. The right Y-axis is the effective diversity and it is plotted as a dashed green line.}
	\label{fig:s3}
\end{figure}

\begin{figure}[H]
	\begin{subfigure}{1\textwidth}
		\includegraphics[width=\textwidth]{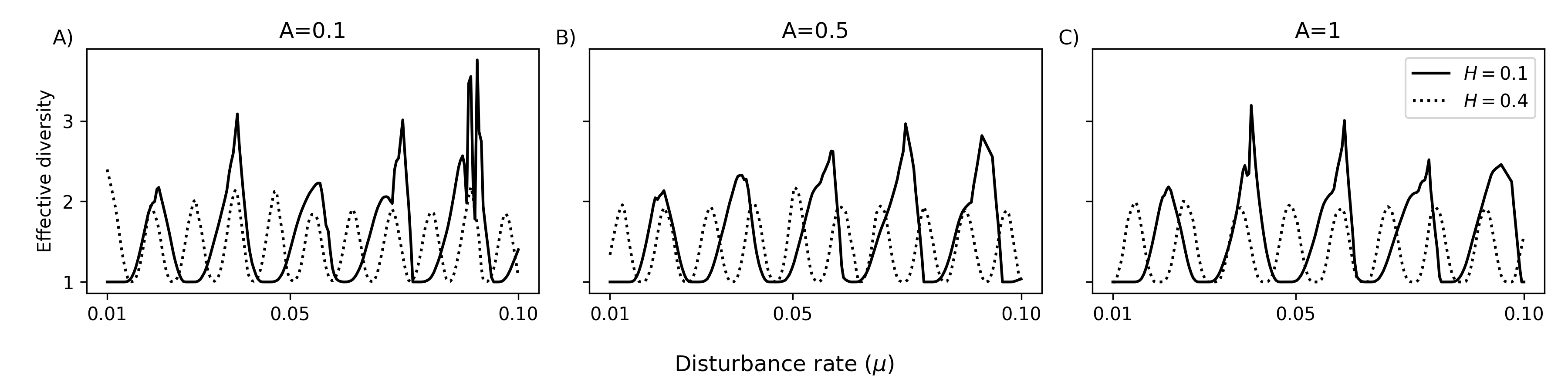}
	\end{subfigure}
	\begin{subfigure}{1\textwidth}
		\includegraphics[width=\textwidth]{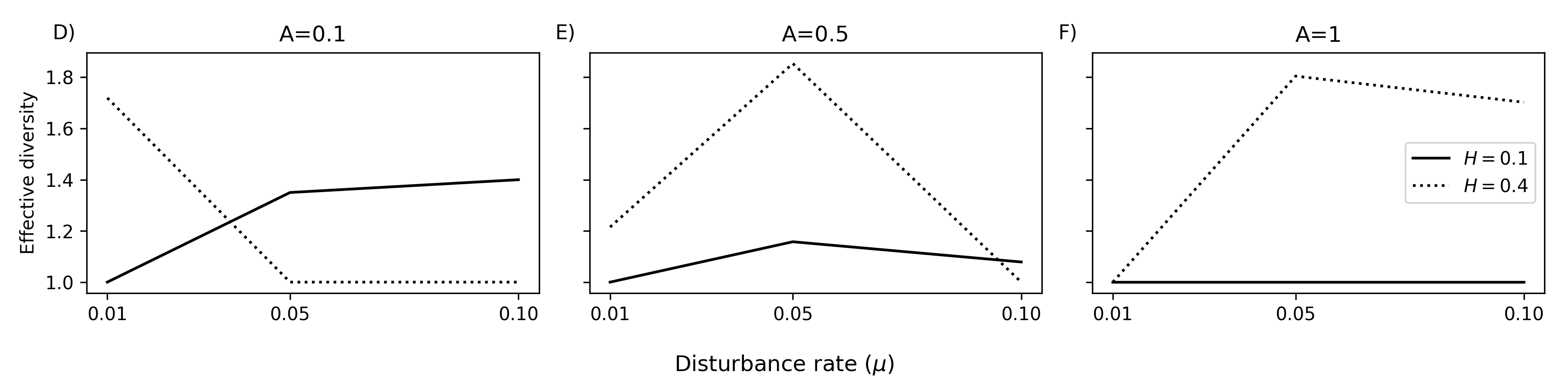}
	\end{subfigure}
	\caption{\textbf{Diversity curves for 50 species.} We depict the relationship between effective diversity and disturbance rate ($\mu$) due to disturbance using 50 species forming a competitive hierarchy, with competing/colonising abilities ($c_i$ = $d_i$) equally spaced between 0 and 1. We sampled the effective diversity over the disturbance gradient continuously (A - C) or discretely (D - F), by sampling three points (0.01, 0.05 and 0.1). We depict the diversity curves for three levels of spatial autocorrelation ($A$), indicated on top of each figure, and two levels of habitat loss ($H$), shown in different line styles.}
	\label{fig:s4}
\end{figure}

\pagebreak
\setstretch{1.24}
\setcounter{section}{0}

\textbf{\LARGE Supplementary file 1: Stability analysis}

\section{General procedure}
Consider the following two-dimensional ODE:
\begin{equation}\begin{split}
		\dv{X}{t} = f(X,Y),
		\\
		\dv{Y}{t} = g(X,Y),
\end{split}\end{equation}
for some positive initial condition.
The positive equilibrium should solve \(f(X_*,Y_*)=g(X_*,Y_*)=0\).
The Jacobian around the positive equilibrium is given by:
\begin{equation}
	J = \begin{pmatrix} \pdv{f}{X} & \pdv{f}{Y} \\ \pdv{g}{X} & \pdv{g}{Y} \end{pmatrix}
\end{equation}
evaluated at \((X_*,Y_*)\).
We then perturb the system around the positive equilibrium:
\begin{equation}
	x \coloneq X-X_*, y\coloneq Y-Y_*
\end{equation}
with \(0\leq x \ll 1\) and \(0\leq y \ll 1\).

Let me denote the eigenvalues of \(J\) by \(\alpha_1\) and \(\alpha_2\), and corresponding eigenvectors by \(\mathbf{U}_1\) and \(\mathbf{U}_2\):
\begin{equation}
	J \mathbf{U}_i = \alpha_i \mathbf{U}_i, \text{ for } i=1,2.
\end{equation}
Suppose \(\alpha_1\neq \alpha_2\) for simplicity (or consider such a case only; otherwise, we need more delicate argument).
Then the solution of the ODE can be in general written as:
\begin{equation}\begin{split}
		x = c_1 \exp(\alpha_1 t) \mathbf{U}_1 + c_2 \exp(\alpha_2 t) \mathbf{U}_2,
		\\
		y = d_1 \exp(\alpha_1 t) \mathbf{U}_1 + d_2 \exp(\alpha_2 t) \mathbf{U}_2,        
\end{split}\end{equation}
for some constants \(c_1,c_2,d_1,d_2\).
For the system to converge (or shrink in size), the maximum real parts of \(\alpha_1\) and \(\alpha_2\) must be negative (i.e., both must be negative).
The eigenvalues satisfy the following equation (characteristic equation):   
\begin{equation}
	\det(\alpha I -J)=0 \Longleftrightarrow \alpha ^2 - \tr(J)\alpha + \det(J)=0. 
\end{equation}
We write \(\tau=\tr(J)\) and \(\delta=\det(J)\).
\begin{compactitem}
	\item When \(\alpha_1\) and \(\alpha_2\) are both real, i.e., when \(\tau^2-4\delta>0\), both of the eigenvalues are negative if and only if \(\delta>0\) and \(\tau<0\).
	\item When \(\alpha_1\) and \(\alpha_2\) are neither real, i.e., when \(\tau^2-4\delta<0\) (which implies that \(\delta>0\)), both of them have negative real parts if and only if \(\tau<0\).
\end{compactitem}
Taken together, the necessary and sufficient condition for the negativity of the maximum of the real parts of the eigenvalues is given by \(\delta>0\) and \(\tau<0\).

\section{Model Description}

We consider the following population dynamics system:

\[
\begin{cases}
	\displaystyle \dv{X}{t} = a(1 - H - X - Y)X - bXY - \mu X \\
	\displaystyle \dv{Y}{t} = b(1 - H - Y)Y - \mu Y
\end{cases}
\]

We assume all parameters \( a, b, \mu > 0 \) and \( 0 < H < 1 \), and restrict our attention to the case where
\[
a(1 - H) - \mu > 0, \quad b(1 - H) - \mu > 0,
\]
which guarantees that both species can thrive in the absence of their counterpart species.

\section{Equilibrium Points}

\subsection*{Boundary Equilibria}

\paragraph{(i) Origin \( E_0 = (0, 0) \):} Clearly, \( \dv{X}{t} = \dv{Y}{t} = 0 \) at the origin.

\paragraph{(ii) \( E_1 = \left(1 - H - \frac{\mu}{a}, 0\right) \):} Set \( Y = 0 \). Then the \( X \)-equation becomes:
\[
\dv{X}{t} = a(1 - H - X)X - \mu X.
\]
Solving \( X[a(1 - H - X) - \mu] = 0 \), we get:
\[
X = 0 \quad \text{or} \quad X = 1 - H - \frac{\mu}{a}.
\]
The second solution is positive under the assumption \( a(1 - H) - \mu > 0 \).

\paragraph{(iii) \( E_2 = \left(0, 1 - H - \frac{\mu}{b}\right) \):} Set \( X = 0 \). Then the \( Y \)-equation becomes:
\[
\dv{Y}{t} = b(1 - H - Y)Y - \mu Y,
\]
yielding:
\[
Y = 0 \quad \text{or} \quad Y = 1 - H - \frac{\mu}{b},
\]
which is positive under the assumption \( b(1 - H) - \mu > 0 \).

\subsection*{Interior Equilibrium \( E_* = (X_*, Y_*) \)}

Solving \( \dv{Y}{t} = 0 \), we obtain:
\[
Y_* = 1 - H - \frac{\mu}{b}.
\]

Substituting \( Y = Y_* \) into the \( X \)-equation and solving \( \dv{X}{t} = 0 \), we get:
\begin{align*}
	a(1 - H - X - Y_*) - b Y_* - \mu &= 0 \\
	aX &= a(1 - H - Y_*) - b Y_* - \mu.
\end{align*}

Using \( Y_* = 1 - H - \frac{\mu}{b} \), we find:
\[
X_* = \frac{1}{a} \left( \frac{a\mu}{b} - b(1 - H - \frac{\mu}{b}) - \mu \right).
\]

\subsection*{Existence Condition of Interior Equilibrium}

From previous derivation:
\[
Y_* = 1 - H - \frac{\mu}{b}, \quad
X_* = \frac{1}{a} \left( \frac{a\mu}{b} - b(1 - H) \right)
\]

Clearly, \( Y_* > 0 \iff b(1 - H) > \mu \), which is assumed.

The condition \( X_* > 0 \) is equivalent to:
\[
\frac{a\mu}{b} > b(1 - H) \iff a\mu > b^2(1 - H)
\]

Hence, the interior equilibrium \( (X_*, Y_*) \) exists in the positive quadrant if and only if:
\[
\boxed{
	\quad a\mu > b^2(1 - H), \quad b(1 - H) > \mu \quad
}
\]

\subsection*{Relation to Stability}

The local stability of the interior equilibrium can only be assessed when it exists. That is, the Jacobian matrix and its eigenvalues are only meaningful at \( (X_*, Y_*) \) if \( X_*, Y_* > 0 \). Therefore, the existence condition \( a\mu > b^2(1 - H) \) is a \textbf{prerequisite} for analyzing local asymptotic stability.

Once existence is ensured, the Jacobian determinant and trace can be evaluated to determine whether the equilibrium is a stable node or focus.

\section{Jacobian Matrix}

Define the functions:
\[
f(X, Y) = a(1 - H - X - Y)X - bXY - \mu X, \quad g(X, Y) = b(1 - H - Y)Y - \mu Y.
\]

The Jacobian matrix is given by:
\[
J = \begin{pmatrix}
	\frac{\partial f}{\partial X} & \frac{\partial f}{\partial Y} \\
	\frac{\partial g}{\partial X} & \frac{\partial g}{\partial Y}
\end{pmatrix},
\]
where:
\begin{align*}
	\frac{\partial f}{\partial X} &= a(1 - H - 2X - Y) - bY - \mu, \\
	\frac{\partial f}{\partial Y} &= -aX - bX = -X(a + b), \\
	\frac{\partial g}{\partial X} &= 0, \\
	\frac{\partial g}{\partial Y} &= b(1 - H - 2Y) - \mu.
\end{align*}

\section{Stability Analysis of Equilibria}

\paragraph{(i) \( E_0 = (0, 0) \):}
\[
J(E_0) = \begin{pmatrix}
	a(1 - H) - \mu & 0 \\
	0 & b(1 - H) - \mu
\end{pmatrix}
\]
Both eigenvalues are positive under the assumptions, so \( E_0 \) is an unstable node.

\paragraph{(ii) \( E_1 = \left(1 - H - \frac{\mu}{a}, 0\right) \):}
\begin{align*}
	\frac{\partial f}{\partial X} &= a(1 - H - 2X) - \mu = -a(1 - H) + \mu, \\
	\frac{\partial g}{\partial Y} &= b(1 - H) - \mu > 0.
\end{align*}
The Jacobian has eigenvalues of opposite signs \( \Rightarrow \) saddle point (unstable).

\paragraph{(iii) \( E_2 = \left(0, 1 - H - \frac{\mu}{b} \right) \):}
\begin{align*}
	\frac{\partial f}{\partial X} &= a\left(\frac{\mu}{b}\right) - \mu = \mu\left(\frac{a}{b} - 1\right), \\
	\frac{\partial g}{\partial Y} &= -b(1 - H) + \mu.
\end{align*}
Again, eigenvalues with opposite signs \( \Rightarrow \) saddle point (unstable).

\paragraph{(iv) \( E_* = (X_*, Y_*) \):} The sign of the trace and determinant of \( J(E_*) \) determines stability. Under certain parameter values satisfying the assumptions, the interior equilibrium can be locally asymptotically stable.

\section*{5. Conclusion}

\begin{itemize}
	\item One interior equilibrium \( E_* \) is existing and also stable, if and only if \(b^2 (1-H) < a \mu < ab(1-H)\).
	\item Three boundary equilibria are saddle points and hence unstable whenever the interior equilibrium is stable.
\end{itemize}

\end{document}